\documentstyle[emulateapj]{article}

\slugcomment{To appear in the Astrophysical Journal}

\begin{document}

\title{Radio Continuum Imaging of the Spiral Galaxy NGC 4258}

\author{Scott D.~Hyman, Daniel Calle\altaffilmark{1}\altaffiltext{1}{Current Address:
Digital Paper, 201 North Union St., Suite 140, Alexandria, VA 22314}}
\affil{Department of Physics, Sweet Briar College, Sweet Briar, VA  24595}
\authoremail{hyman@sbc.edu}
\authoremail{dcalle@geocities.com}

\author{Kurt W.~Weiler, Christina K.~Lacey\altaffilmark{2,3}\altaffiltext{2}{National Research Council 
Post-Doctoral Fellow}\altaffiltext{3}
{Current Address: Department of Physics and Astronomy, University of South Carolina, Columbia, SC 29208}}
\affil{Naval Research Laboratory, Code 7213,
Washington, DC 20375-5320}
\authoremail{weiler@rsd.nrl.navy.mil}
\authoremail{lacey@rsd.nrl.navy.mil}

\author{Schuyler D.~Van Dyk}
\affil{IPAC/Caltech, 100-22, Pasadena, CA  91125}
\authoremail{vandyk@ipac.caltech.edu}

\and

\author{Richard Sramek}
\affil{National Radio Astronomy Observatory, P.O. Box 0, Socorro, NM 87801}
\authoremail{dsramek@nrao.edu}

\begin{abstract}

We analyze $3{\farcs}5$ resolution, high sensitivity radio continuum images 
of the nearby spiral galaxy NGC 4258 at 6 and 20 cm derived from
multiple observations used to monitor the radio supernova SN 1981K (Van Dyk et al.~1992,
[ApJ, 396, 1995]). Seven bright H~II region
and five supernova remnant candidates are identified. Extinctions to the H~II regions
are estimated for the first time from a comparison of radio flux
densities to new optical fluxes derived from H$\alpha$ observations
by Dutil \& Roy (1999, [ApJ, 516, 62]). The bright end of the H~II region luminosity 
function is established at
each wavelength. The luminosity functions are best fit by power
laws consistent with the shape of previously published radio and optical
luminosity functions for a number of galaxies.  The supernova remnants are all about 2--3
times the radio luminosity of Cas A.  In addition, the galaxy's nucleus is shown to have been 
variable over the SN 1981K monitoring period.
The spectral index ($\alpha$) distribution of the anomalous radio arms is investigated and found to be
relatively uniform at $\alpha = -0.65 \pm 0.10$. 
 
\end{abstract}


\section{Introduction}

Radio continuum emission from normal galaxies is a combination of both thermal and nonthermal 
discrete emission from individual sources and diffuse emission from the galactic disk.
The individual thermal emitters are assumed to be free-free emitting H~II regions, which affect 
the interstellar medium (ISM) through the ionizing radiation of young stars, and the
nonthermal emitters are generally thought primarily to be supernova remnants (SNRs), which
also profoundly affect the ISM by injecting enormous amounts of kinetic energy
and enriched material.  SNRs are the leading candidates for the production of cosmic rays and 
contributors to the relativistic ISM.  Thus, studies of H~II regions and SNRs provide important 
information on the beginning and endpoints of stellar evolution and reveal much about
the evolution of the ISM and, ultimately, the evolution of galaxies.

Analyses of extragalactic SNRs and H~II regions at radio wavelengths are not hampered by 
the distance ambiguities of Galactic samples and extinction by dust and gas.  A relatively 
well-defined samples of SNR and H~II region distributions in external galaxies can be used to 
yield an unbiased statistical picture of group properties, which can yield important information 
on star formation and the role of supernova (SN) and SNR shock waves in
determining the energetic and dynamical evolution of the ISM, and, perhaps, even in triggering 
new epochs of star formation.  Radio observations of H~II regions are nearly a direct measure of
the ionizing luminosity of the embedded stars.  A comparison with optical line luminosities can 
result in estimates of the line-of-sight extinction, providing an alternative to hydrogen line 
ratio methods alone, and can make it possible to trace the distribution of dust and
the nature of extinction gradients in other galaxies.  Extinction measurements can also help 
determine the amount of dust internal to individual H~II regions, which are essential for 
accurate estimates of atomic and molecular abundances and chemical gradients within galaxies.
Comparisons between thermal radio and far-infrared properties can lead to a better understanding 
of the young stars that power H~II regions and their role in the galactic-scale, far-infrared
emission properties. 

Due to limited resolution and sensitivity, large samples of discrete radio sources in external 
galaxies have only been observed in the Local Group (e.g., in M31, Braun \& Walterbos 1993; in 
M33, Viallefond \& Goss 1986, Duric et al.~1993, Duric et al.~1995, and Gordon et al.~1999;
in the LMC and SMC, Mills 1983 and Mathewson et al.~1983), in intense starburst galaxies 
(e.g., in M82, Kronberg, Bierman, \& Schwab 1985; in NGC 253, Ulvestad \& Antonucci 1997), in
M51 (van der Hulst et al.~1988), in M81 (Kaufman et al.~1987), and in
NGC 6946 (Lacey, Duric, \& Goss 1997, Hyman et al.~2000).  As part of a larger study of the
continuum emission from nearby galaxies, we consider in this paper the case of NGC 4258.

NGC 4258 (M106) is a large ($18{\farcm}6 \times 7{\farcm}2$; NASA/IPAC Extragalactic Database, NED), 
nearby (7.3 Mpc; Herrnstein et al.~1997) spiral galaxy 
with high inclination 
($i = 71\arcdeg$, Sandage \& Tammann 1981), at high galactic latitude
($b = 68{\fdg}8$), with correspondingly low foreground extinction ($A_B = 0.00$; NED). 
The galaxy is also notable for its LINER/Seyfert nuclear properties, for the
detection of a very bright radio continuum nuclear source (Turner \& Ho 1994),
for the discovery of H$_{2}$0 masers and for the possibility of a supermassive 
black hole at the center (Miyoshi et al.~1995; Greenhill et al.~1995).
H$\alpha$ imaging by Courtes et al.~(1993) and Dutil 
\& Roy (1999) show a number of very bright H~II regions along the inner spiral arms of the galaxy.

NGC 4258 has been previously studied at numerous radio wavelengths. van der Kruit, Oort, \& 
Mathewson (1972)
observed the galaxy at 20 cm with the Westerbork Synthesis Radio Telescope (WSRT) at 
24$\arcsec$ $\times$ 32$\arcsec$ resolution, revealing very bright and extensive ``anomalous'' 
radio arms in the east-west direction.  de Bruyn (1977), with the WSRT at 6 and 49 cm with 
resolution 7$\arcsec$ $\times$ 9$\arcsec$ and 56$\arcsec$ $\times$ 76$\arcsec$, respectively,
found a nonthermal spectral index for the anomalous arms.
van Albada \& van der Hulst (1982), with the WSRT ($6{\farcs}5$ beam) and the Very Large
Array (VLA\footnote{The Very Large Array (VLA) is a telescope of the National Radio Astronomy 
Observatory (NRAO) which is operated by Associated Universities, Inc., under a cooperative 
agreement with the National Science Foundation.}; $11{\farcs}25$ beam), both at 20 cm, 
found that the anomalous arms are multiply branched. 
Krause, Beck, \& Klein (1984), with the Effelsberg 100-m telescope at 6.3 and 2.8 cm 
($\sim$1$\arcmin$ resolution), detected no significant spectral index variations along the 
anomalous arms.  Hummel, Krause, 
\& Lesch (1989) conclude, based on their analysis of 
the polarized emission using the VLA at 6 and 20 cm (14$\arcsec$ resolution) and Effelsberg at 
1.2 cm (40$\arcsec$ resolution), that the anomalous arms lie in the plane of NGC 4258. 

Although the anomalous arms coincide with faint H$\alpha$ emission (Courtes \& Cruvellier 1961; 
Courtes et al.~1993; Dutil \& Roy 1999), they were originally thought to be out of the plane of 
the optical galaxy (Courtes \& Cruvellier 1961; van der Kruit et al.~1972),
due to the absence of H~II regions along them and other considerations. The radio arms have been
modeled as arising from the expulsion of gas from the nucleus either in a single event 
(van der Kruit et al.~1972) or by continuous ejection in a jet (Martin et al.~1989; Ford et 
al.~1986; Cecil, Wilson, \& Tully 1992; Pietsch et al.~1994).  However, the CO(1--0) observations
by Krause et al.~(1990) revealed a large quantity of molecular gas along the radio arms which 
would have required an unreasonably large energy output in the single event model.  
Recently, Cecil et al.~(2000), with VLA 20 cm observations at
$1{\farcs}3$ resolution and high-resolution optical data, conclude that the 
arms arise from changing jet activity at the nucleus.
Cox \& Downes (1996), 
from CO(2--1) observations of the inner parts of the anomalous arms, 
offer an alternative explanation, concluding that the arms arise from gas flows and shocks 
due to a galactic bar potential within the disk of NGC 4258.  

The previous radio studies of NGC 4258 focused mostly on the analysis of the 
anomalous arms and not on the discrete sources in the galaxy, other than the
nucleus.
We present here our high resolution ($3{\farcs}5$) and high sensitivity 
($\sim$25 $\mu$Jy beam$^{-1}$) observations, which result in a detailed analysis of the 
H~II regions and SNRs.  Although our study does not concentrate on the nature
of the anomalous arms, our radio maps 
show significantly smaller structures in the anomalous arms and smaller-scale 
spectral index variations within them.  Optical-to-radio comparisons are used to estimate the 
visual extinctions toward the H~II regions.  Also, the bright end of the H~II region luminosity 
function is produced, and the star formation properties of the brightest H~II regions are also 
presented.  The SNRs are shown to be very luminous in the radio, and the 
nucleus appears to exhibit variability over more than a decade.

\section{Radio Observations}

The Type II radio SN 1981K in NGC 4258 was monitored with radio continuum VLA observations 
at 6 and 20 cm 
from 1985 March through 1990 May in all four array configurations. Details of the data 
reduction for these observations and the results of the analysis for SN~1981K are described 
in Weiler et al.~(1986) and Van Dyk et al.~(1992).
We have used the AIPS task DBCON to combine the observations at each frequency from 1985 
March 30 through 1990 May 29, and from additional SN~1981K monitoring observations
made on 1991 July 29, 1991 October 31, and 1994 May 03 in A and B configurations.
The integrated times-on-source for the concatenated uv databases are 16.6 hr at 6 cm and 
10.6 hr at 20 cm. The uv data were Fourier-transformed and beam-deconvolved using two 
different procedures in AIPS.  First, the task IMAGR was used.  Images with equal resolutions 
of $3{\farcs}4 \times 3{\farcs}3$ at both wavelengths were generated by appropriately 
weighting the data with UVTAPER.  These images were used to analyze most of the discrete 
sources in and background sources surrounding NGC 4258. However, the anomalous radio arms
have a very speckled appearance in these images, due to cleaning of the very 
extended emission with a narrow beam.  For that reason, the task VTESS, running
the Maximum Entropy Method (MEM), was used.  While the CLEAN algorithm in IMAGR is more 
reliable for imaging discrete sources, MEM is more suitable for imaging extended emission.
Zero-spacing fluxes of 0.40 Jy (Krause et al.~1984) and 0.82 Jy (Hummel et al.~1989), 
corresponding to the integrated flux densities at 6 and 20 cm, respectively, were input to 
VTESS to correct for missing short spacings in the uv coverage. The MEM images of NGC 4258 
are shown in Figures 1a and b. They have nearly the same resolution ($3{\farcs}5 \times 3{\farcs}3$)
and sensitivity (20 $\mu$Jy beam$^{-1}$ and 30 $\mu$Jy beam$^{-1}$ at 6 and 20 cm, respectively) 
as the images made with IMAGR (not shown).  The labels in Figure 1a refer to discrete sources 
discussed in \S 4.1.1 and \S 4.1.2 below. 

Corrections for primary beam attenuation have been applied to both the CLEAN and MEM images,
since the anomalous arms extend $\sim$8$\arcmin$ east-west and the prominent inner north-south 
spiral arms extend $\sim$6$\arcmin$, well within the 30$\arcmin$ FWHM primary beam at 20 cm, but
less so, within the $\sim$9$\arcmin$ (FWHM) primary beam at 6 cm.  Since the field center 
of the observations is at the location of SN~1981K ($\alpha = 12^h 18^m 59.42^s$,
$\delta = +47\arcdeg 19\arcmin 31{\farcs}0$; J2000), the 50 MHz bandwidth at both wavelengths 
did not result in significant bandwidth smearing for sources along the northern arm. However, 
smearing is significant at 20 cm for sources in the southern arm, where the reduction in the peak 
flux density of a point source is as much as a factor of two (Bridle \& Schwab 1988).  While the 
integrated flux density of a bandwidth-smeared bright source, such as the prominent source at the 
end of the southern spiral arm, should not be affected, fainter sources could be smeared into the 
surrounding arm emission and noise, and hence, be difficult to detect. 
The detection of faint radio sources in the southern arm is discussed further in the next section.

\section{Source Detection and Measurement}

Because of the complexity of the radio emission from NGC 4258, it was often difficult to resolve
discrete sources from background disk and anomalous arm emission. Therefore, only the brightest 
sources (nos.~1--9; Figure 1a), a few fainter sources in the southern arm (nos.~10-12; Figure 3), 
one background source at the tip of the southeastern radio arm, and nine background sources in located outside
the field of Figures 1a and b were chosen for analysis. The CLEAN maps were used to measure the properties
of sources 1--9, the nucleus, and the background sources.  The MEM images were used to measure 
sources 10--12 and to generate a spectral index map of the anomalous radio arms
(see \S 4.4.2).  In addition to our radio data, Y.~Dutil and J.-R.~Roy kindly provided to us
narrow-band, continuum-subtracted H$\alpha$ imaging data for NGC 4258 (see 
Dutil \& Roy 1999 for details).  Briefly, the H$\alpha$ images were acquired
with the 1.6 m telescope at Mont Megantic Observatory, the
two satellite [N~II] lines were excluded using a 10 \AA-wide 
filter, the seeing was $\sim$2-3$\arcsec$, and the optical flux calibration 
uncertainty is $\sim$20$\%$.  We registered the radio and H$\alpha$ images, 
and show the H$\alpha$ map overlaid on the 20 cm MEM greyscale image in Figure 2.

The positions of most of the brightest radio sources have optical counterparts with peak pixel 
positions that agree to within $1{\farcs}5$.  The brightest portions of the anomalous arms also 
coincide with diffuse optical emission, although significantly more structure
is apparent in the radio images (see \S 4.4.1).  In contrast to the radio images in Figures 1a and b, 
the optical-radio comparison in Figure 2 shows significant H$\alpha$ emission and many discrete 
sources in the southern arm, besides the very bright source 9.  
Because of the confusion of faint sources with background emission, we have used the MEM images 
to identify sources in the southern arm.  An MEM-derived contour map of the 20 cm radio emission 
from the southern arm and surrounding regions is shown in Figure 3.  The radio map 
shows faint emission coincident with several of the discrete H$\alpha$ sources in the southern arm. 
  
The flux densities of the radio sources were measured using the same technique as that used for 
NGC 6946 (Lacey et al.~1997; Hyman et al.~2000). First, the total flux was measured within an
aperture including the source and an annulus outside the aperture measuring the background.
The average background was computed and subtracted from the flux within the source aperture to 
obtain the source flux density. Note that source 4 appears to have two components in the 20 cm 
image (Figure 1a).  However, since the corresponding source on the 6 cm and H$\alpha$ images does not
show any structure and covers the same area as the 20 cm source, we have simply measured the flux 
using one inclusive aperture.  Fluxes were measured for each of the optical counterparts to radio
sources using the same method.  Since the resolution of the radio maps is similar to the
seeing of the optical observations, identical source apertures were used. 

\section{Discussion}

\subsection{Source Criteria}

The thermal radio emission from H~II regions is identifiable by its flat
spectral index, $\alpha$ (where $S \propto \nu^{+\alpha}$), with a nominal 
value $\alpha = -0.1$.  For nonthermal emission from, e.g., SNe, SNRs, and 
optically thin active galactic nuclei, $\alpha$ is steeper, in the approximate 
range $-0.4$ to $-1.0$.  Unfortunately, a nonthermal plerionic SNR, which also 
has a flat spectrum ($-0.3 \le \alpha \le -0.1$; Weiler \& Panagia 1978), 
might be confused with an H~II region by this criterion.  However, 
extrapolating from the small number of plerions compared to the
shell-type remnants (Srinivasan et al.~1984), it is likely that we would 
correctly identify the SNRs in NGC 4258 down to the surface brightness limit,
with only one or two plerions possibly misidentified.

\subsubsection{Discrete Sources in NGC 4258}

In Table~1 we present the positions, flux densities, and spectral indices for
12 discrete radio sources, including the galactic nucleus. The position listed
for each source corresponds to the 
centroid at 20 cm.  SN 1981K is not included in the table. 
Although the uncertainties in the spectral indices are appreciable, we employ
the criteria above to discriminate between thermal and nonthermal sources.
Of the 13 sources, sources 1--6 and 9 are considered
thermal, with $\alpha$ between +0.05 and $-0.25$, and are, 
therefore, classified as H~II region candidates.
Two bright sources (7 and 8) and the faint southern arm sources
(10, 11, and 12) have nonthermal $\alpha$ between $-0.37$ and $-0.82$, and
are classified as SNR candidates. 

We have identified and measured optical counterparts on the H$\alpha$ image 
to the 13 radio sources in Table 1, except for two SNR candidates (sources 7 
and 8) and source 6, which is blended with source 5 on the H$\alpha$ map (see 
below).  The optical fluxes, as well as estimates of visual extinction toward 
the H~II regions, are listed in the table (see also \S 4.1.5).  The nucleus, source
2, and the nonthermal sources 11 and 12 are coincident with, or
are near, four X-ray sources (X18, X16, X19, and X14, respectively;
Vogler \& Pietsch 1999). The X-ray counterparts to sources 11 and 12 are 
indicated by boxes in Figure 3. 

Note from Figure 1a that one optical source covers an area that includes both 
radio sources 5 and 6; therefore, a separate aperture including
both sources was used to measure the optical flux and the flux density of the
combined pair of radio sources (source ``5+6'' in Table~1). Also, while an 
optical source is located $4{\farcs}2$ from the nonthermal source 7 (i.e., 
somewhat more than one synthesized half-power beam width), we have not included
its flux in the table, since typically the radio and optical source positional
uncertainties are less than $\sim$$1\arcsec$. 

\subsubsection{Background Sources}

Based on estimates of the background source counts with flux densities above
100 $\mu$Jy at 21 cm (Oort 1987), we expect $\sim$4 background sources to 
occur in the field.  It is therefore possible that extragalactic radio sources 
with nonthermal spectra could be confused with our SNR candidates, or that a 
flat spectrum galaxy nucleus could be confused with our H~II region candidates.
However, since all thermal sources and all three southern arm SNR candidates
also have H$\alpha$ counterparts, these sources are less likely to
be background objects. Although the nonthermal source 7 does not have a 
clear optical counterpart, its location in the northern arm near the central 
region of the galaxy is supportive of an identification as a SNR.  The 
nonthermal source 8 could possibly be a background object, since it is
has no optical counterpart and is located in the interarm region. 

One notable, very bright, nonthermal radio source within the field of Figures 1a
and b (source 15), located east of the nucleus at the tip of the southeastern 
radio arm, was rejected as a SNR candidate because of its possible ``head-tail''
structure. The 20 cm position of the source's bright (bandwidth smeared) 
``head'' component is $\alpha = 12^h 19^m 18{\fs}92$,
$\delta = +47\arcdeg 19\arcmin 07{\farcs}6$ (J2000).
The bifurcated radio ``tail'' extends northward. At 6 cm, only the northeastern
part of the tail is clearly visible above the noise. The head has a 20 cm flux 
density of $2.09 \pm 0.23$ mJy and a nonthermal spectral index 
$\alpha = -0.69 \pm 0.15$.  Hummel et al.~(1989) and van Albada \& van der 
Hulst~(1982) also detected this source in their lower-resolution maps and 
suggested that, given its extreme brightness, nonthermal spectral index, and 
possible complex structure, it is a background radio galaxy rather than a 
source associated with NGC 4258.  From its morphology evident in our 
higher-resolution maps, we agree with that conclusion.

In Table~2 we list source 15, described above, two clusters of sources 
(16 and 17), and two additional sources (13 and 14), all
lying within the 20\% attenuation level of the primary beam at 6 cm, but, with
the exception of source 15, outside the field shown in Figures 1a and b.  These
sources have no optical counterparts.  Nine of the ten sources listed in the
table have nonthermal spectral indices ranging from $\alpha = -0.7$ to $-1.1$. 
The cluster of sources 16a--d corresponds to source 6 in Hummel et al.~(1989),
which they list as a ``triple,'' and the cluster of sources 17a--c corresponds
to their source 2, listed as a ``double.''  Hummel et al.~(1989) and van Albada \& van der
Hulst (1982) suggest that the clusters constitute two radio galaxies. Indeed, 
the source cluster 16a--d may consist of a flat spectrum 
($\alpha = -0.15 \pm 0.11$) nuclear source (16b) with nonthermal radio lobes 
to the north (16a) and south (16c, d). Similarly, the sources 17a and 17c may 
be the lobes of a radio galaxy, with the nucleus located at the position of 
source 17b.  While additional sources are visible within the larger 
field-of-view at 20 cm, they are extremely bandwidth smeared; hence, no 
measurements were attempted for these sources, especially since no information 
is available for them at 6 cm.

\subsubsection{H~II Region Star Formation Properties}

The seven radio H~II regions are among the brightest optically in NGC 4258,
with $L_{{\rm H}\alpha} \gtrsim 10^{38}$ erg s$^{-1}$ (Courtes et al.~1993).
We have calculated the star formation rate, SFR, for massive stars from their
thermal radio spectral luminosity, $L_{\rm T}$, using Eq.~23 of Condon~(1992):
\begin{equation}
L_{\rm T} \sim 5.5 \times 10^{20}(\nu^{-0.1})(SFR)
\end{equation}
where $L_{\rm T}$ is in W Hz$^{-1}$, $\nu$ is in GHz, and SFR is in M$_{\sun}$ yr$^{-1}$. Condon~(1992) derives 
this relation in the context of the global star formation properties for normal galaxies using the ``extended''
Miller \& Scalo~(1979) 
initial mass function $\psi(M) \propto M^{-5/2}$, for stellar masses above 5 M$_{\sun}$ and below $\sim$100 M$_{\sun}$.
The relatively dust-free $L_{\rm T}$ was
determined at 6 cm assuming a distance of 7.3 Mpc to
NGC 4258 (Herrnstein et al.~1997).  
For H~II region candidates with
$-0.25 < \alpha < -0.1$ (sources 3, 4, 5, and 9), 
the uncertainties in $\alpha$ are significant and,
therefore, we cannot rule out a nonthermal component. 
A correction was first applied for residual nonthermal contamination
which was, presumably, not fully removed in the
background-subtraction procedure discussed in Section 3.  
A nonthermal spectral index of $\alpha_{\rm nt} = -0.8$ was
assumed for the contamination. This correction is at the $\sim$1$\sigma$ level,
except for source 9, for which the flux density is
reduced by 2$\sigma$ and 3.5$\sigma$ at 6 and 20 cm, respectively. 
We present in Table 3 for each H~II region the corrected thermal flux densities 
(S$^{\rm th}_{6,20}$), $L_{\rm T}$, the SFR, the production rate of Lyman 
continuum photons ($N_{\rm UV}$, in s$^{-1}$),
and the radio supernova rate ($\nu_{\rm SN}$, in yr$^{-1}$). 
$N_{\rm UV}$ and $\nu_{\rm SN}$ were calculated using Eqs.~24 and
20 from Condon~(1992):
\begin{equation}
\nu_{\rm SN} \sim 0.041 (SFR)
\end{equation}
\begin{equation}
N_{\rm UV} \sim 3.5 \times 10^{53} (SFR),
\end{equation}
where the latter equation assumes that dust absorption of Lyman continuum 
photons is negligible.

\subsubsection{H~II Region Luminosity Functions}

H~II region luminosity functions, i.e., the number distribution of source
luminosities, can reflect the mass distribution of the parent molecular clouds 
and the efficiencies for conversion of molecular gas during star formation.  
Considerable foreground extinction impedes H$\alpha$ studies of H~II regions in
the Galaxy. 
Extragalactic luminosity functions are important, since the lack of known 
distances, sample incompleteness, and selection biases in the Galactic sample 
prevent accurate luminosity functions from being calculated.
Previous radio observations of extragalactic H~II regions, while not affected by extinction
and relative distance ambiguities, usually suffer from limited resolution and sensitivity and the
related difficulties in nonthermal background subtraction.  Thus far, radio
luminosity functions of H~II regions are available for non-Local Group galaxies for only M81
(Kaufman et al. 1987), M51 (van der Hulst et al.~1988), 
NGC 6946 (Hyman et al.~2000), and certain starburst galaxies. 

Since we have detected only seven bright radio H~II regions in NGC 4258, we 
can only plot the brightest end of the luminosity function. 
The cumulative radio luminosity functions,
i.e., $N(S)$ (where $N(S)$ is the number of sources with
 radio flux densities $\ge S$) {\it versus\/} $S$, for the seven H~II region
candidates in Table~1 are shown in Figure~4 for both 6 cm (diamonds) and 20 cm (circles) wavelengths.
The luminosity functions have nearly identical shapes
and power-law indices $\beta$
($N(S) \propto S^{\beta}$), with $\beta = -1.3 \pm 0.2$ 
at both wavelengths. 

The values we obtain for $\beta$ are similar to those found in earlier 
studies. Recently, we reported values of $\beta = -1.2
\pm 0.2$ at 6 cm and $\beta = -1.3 \pm 0.2$ at 20 cm for the bright end of the 
luminosity function formed from 43 radio
H~II region candidates detected in NGC 6946 (Hyman et al.~2000; Lacey et al.~1997).
Viallefond \& Goss (1986) found $\beta = -1.18 \pm 0.14$ for
the H~II region 20 cm luminosity function of M33.
Kaufman et al.~(1987) and van der Hulst et al. (1988) have fit
their 20 cm luminosity functions for H~II regions in M81 and M51 and obtained
$\beta = -1.8 \pm 0.4$ and $\beta \sim -2$, respectively, for the
brightest sources.
Israel (1980) obtained $\beta$ values ranging from $-1.1$
to $-5.5$ for the high end of the radio luminosity functions for 14 late-type 
galaxies, with a steeper value of $-1.7$ for NGC 4258. 
The difference in values probably arises from resolution differences and significant 
background contamination.
We conclude that the index determined from our high end
luminosity function
is likely more reliable. We can clearly distinguish the H~II regions in the northern arm from each
other and can remove background contamination more accurately.  

Courtes et al.~(1993) have derived the H$\alpha$ luminosity function from measurements of 137
H~II regions in NGC 4258.  However, they could not fit a single power law 
to the bright end because of a break in the slope  
seen at 10$^{38.7}$ ergs s$^{-1}$ (this break in a number of galaxies may be due to
an unknown mechanism that inhibits the formation of supergiant H~II regions 
[Kennicutt, Edgar, \& Hodge 1989)], to a transition from ionization- to 
density-bounded regions [Rozas, Beckman, \& Knapen 1996], or to a variable, 
rather than constant, star formation evolution [Feinstein 1997]).
From Fig.~4 of Courtes et al.~(1993) for $L \gtrsim 10^{38}$ erg s$^{-1}$, we estimate that the
slope $\beta \simeq -1$, somewhat shallower than the radio-determined slope.

\subsubsection{H~II Region Extinction}

From our radio and optical measurements we have estimated the
extinction at H$\alpha$, $A_\alpha$, toward the H~II regions listed in Table~1,
using  Eq.~2 of Kaufman et al. (1987):
\begin{equation}
A_{\alpha} = 2.5 \log_{10}[C(\nu,T_e)S_{\nu}/S_{H\alpha}]
\end{equation}
where the electron temperature $T_e$ is assumed to be 10$^4$ K, and 
$C(\nu,T_e)= 8.01 \times
10^{-13}$erg cm$^{-2}$ s$^{-1}$ mJy$^{-1}$ at $\nu$=1.465 GHz
and $C(\nu,T_e)= 9.05 \times
10^{-13}$erg cm$^{-2}$ s$^{-1}$ mJy$^{-1}$ at $\nu$=4.885 GHz (Gebel~1968).

We obtain mean values of $2.4 \pm 1.0$ at 6 cm and $2.5 \pm 1.0$ at 20 cm for the 
visual extinctions, $A_{V}$, after applying a standard 
galactic reddening law, $A_{V}$ = 1.28$A_{\alpha}$ (Miller \& Mathews 1972). 
Due to NGC 4258's high galactic latitude, the Galactic foreground extinction 
is negligible ($A_{B}$ = 0.00; NED).
We find no correlation between $A_{V}$ and radio flux density, which implies 
that the optical and radio H~II region luminosity functions should have 
similar shapes.  Such a correlation was also
not found for H~II regions in NGC 6946 (Hyman et al.~2000), M81 (Kaufman et 
al.~1987), and M51 (van der Hulst et al.~1988).

The highest extinction is found for
the combination of sources 5 and 6, which are located close to the center of 
NGC 4258, where more dust might be expected. We also
have obtained an estimate of $\sim$5.5 mag for the extinction toward the flat spectrum nucleus which is comparable to that for
sources 5 and 6, and significantly greater than for those with lower extinctions located further from the center. 
Recently, Chary et al.~(2000) from high-resolution infrared observations 
estimate a visual extinction of 18 mag for
the toward the compact (size $< 7$ pc) source at the center of NGC 4258. 

The model by Sarazin (1976) predicts that internal dust in H~II regions would 
give rise to an extinction gradient with galactocentric radius,
related to metal abundance gradients. While a shallow metal abundance gradient (Dutil \& Roy~1999) and
a moderate extinction gradient (Dutil 1998) are found from optical data alone 
for NGC 4258, our sample of extinction measurements is too limited to show a
clear trend.  However, an extinction gradient was also not found for NGC 6946 (Belley 
\& Roy 1992; Hyman et al.~2000), M81 (Kaufman et al.~1987), and M51
(van der Hulst et al.~1988).

Our mean $A_{V}$ values are about 1 mag higher than the mean value of 1.3 mag
obtained for 122
H~II regions in NGC 4258 using Balmer decrements (Dutil \& Roy 1999; Dutil 1998).
Their extinction values, $A_{V}$(Bal), for sources corresponding to our H~II region candidates
are listed in Table~1.
The largest differences (from $\sim$0.5 to $\sim$2 mag) are for sources 1, 9, and the combination of 
5 and 6, and the smallest differences ($\lesssim$0.2 mag) are for sources 2, 3, and 4.
In other galaxies, such as NGC 6946 (Hyman et al.~2000), M33 (Viallefond \& Goss 1986), 
M51 (van der Hulst et al.~1988), and M81
(Kaufman et al.~1987), the radio-determined extinction is also higher than the
Balmer decrement extinction. This discrepancy has been attributed to
patchy dust internal and external to H~II regions, to aperture differences
between the radio and optical measurements, or to the variation of line ratios over the 
spatial extent of the source. 

A comparison of extinction estimates using the radio-to-optical method for other galaxies shows that 
our mean value of 2.5 $\pm$ 1.0 for NGC 4258 
is similar to that of $2.2 \pm 0.7$ for M51 (van der Hulst et al.~1988), but is
greater than $1.0 \pm 1.2$ for NGC 6946
(Hyman et al.~2000) and $1.0 \pm 0.4$ for M81 (Kaufman et al.~1987).

\subsection{Supernova Remnants}

The SNR candidates, sources 7, 8, and 10--12, for a distance of 7.3 Mpc, have
spectral luminosities at 6 cm of 1.6, 1.4, 1.2, 1.7, and $2.0 \times 10^{22}$
erg s$^{-1}$ Hz$^{-1}$, respectively.  These luminosities are all $\sim$2--3 
times the 6 cm luminosity of Cas A, and are comparable to the extraordinary and
very luminous SNR in NGC 6946 (Van Dyk et al.~1994).  Samples of extragalactic 
SNRs, not subject to the extinction and source confusion problems of Galactic SNRs, 
although growing, are still quite small, and these very bright examples are 
particularly intringuing, since they may the remnants of very massive single stars
(Dunne, Gruendl, \& Chu 2000) or
may be multiple interacting SNRs (Blair, Fesen, \& Schlegel 2001).  These bright sources
are deserving of further observational attention.  No fainter likely radio 
SNRs are detected in NGC 4258.

Cecil et al.~(2000) claim that what is our source 7 is a putative northern radio 
``hot spot'' created by the nuclear jet impacting the ambient medium.  Although we
clearly agree that the source is nonthermal, Cecil et al.~(2000) measure a size of 14\arcsec\ 
at $1{\farcs}3$ resolution at 20 cm; at the same frequency we measure at 4\arcsec\
resolution a size of at most 5\arcsec\ (FWHM).  We find the size difference hard to
reconcile and speculate on how much background emission was included in their
measurement.  We therefore call into question its identification as a hot spot of
the jet outflow.   Similarly, Cecil et al.~(2000) isolate a corresponding southern hot spot;
this source is even more embedded in the background, and Figure 1a shows it to
be complex in morphology.  Therefore, we have not included it in our detected source 
list.  This source is most certainly nonthermal as well, with $\alpha$ between $-0.7$
and $-0.4$, but, again, the diffuse background is too bright to make an accurate
estimate.

\subsection{Nucleus}

In our concatenated dataset, the nucleus of NGC 4258 has a relatively flat spectrum, 
$\alpha = -0.24 \pm 0.10$, between 6 and 20 cm.  Turner \& Ho (1994) found from observations 
in 1982 and 1983 an inverted spectrum, $\alpha = +0.4$, between 2 and 6 cm, 
characteristic of an opaque synchrotron source.  We undertook mapping the 
individual observations at each frequency made only in the A-configuration (to image
the nucleus at the highest resolution and to minimize background contamination), to look 
for possible variability in the radio emission from the nucleus.  We measured the integrated
flux density of the source using the tasks IMFIT and IMEAN.  (Uncertainties in the measurements
are derived in a similar fashion to those for SN 1981K by Van Dyk et al.~[1992].)
The sampling in time is 
not very frequent, but Figure 5 indicates that during the monitoring of SN 1981K, the
flux density varied by as much as 47\% at both 6 and 20 cm, {\it possibly\/}
undergoing outburst activity.  The spectral index varied from $-0.18$ to $-0.41$ during this
interval.  To our knowledge, this is the first representation of the
variability of the nucleus of NGC 4258.  Variability of this kind is also exhibited by the 
LINER nucleus of M81 (Ho et al.~1999), albeit in much greater detail.
Such behavior is consistent with the picture of a supermassive black hole being 
responsible for the activity in the radio from the nuclei of both galaxies.

\subsection{The Anomalous Radio Arms}

\subsubsection{Description}

Our high resolution images of NGC 4258 at 6 and 20 cm (see Figures 1a and b) show
greater detail in the anomalous radio arms not visible in the previous lower 
resolution, lower sensitivity maps presented elsewhere (e.g., Hummel et 
al.~1989; van Albada \& van der Hulst 1982), in particular, a large number of
knots and filaments along both arms.  As noted by previous studies, the general
appearance of the northwestern and southeastern arms is quite different.
Many of the individual sources listed by Turner \& Ho (1994) along the arms
are actually just the peak emission in the arms of the various knots.

The brightness of the northwestern anomalous arm decreases first to the north of the central region of NGC 4258 and then brightens again by
$\sim$2.5 times at both wavelengths, to a maximum of 0.54 mJy beam$^{-1}$ and 
1.2 mJy beam$^{-1}$ at 6 and 20 cm, respectively. 
The half-power width of this northwestern arm at the position of maximum flux density before bifurcation
is $7{\farcs}2 \pm 1\arcsec$ 
after correcting for the beamsize broadening.
The brightening and the subsequent
bifurcation of the northwestern anomalous arm about $\sim$1$\arcmin$ from the nucleus is apparent on even very low
resolution maps.  The northern and southern subsections of this arm after 
bifurcation also dim and then brighten considerably
after bifurcation. While the northern section widens and further fragments into several bright filamentary
structures, the southern section remains narrow and brightens into only one main filament.

In contrast to the northwestern anomalous radio arm, the southeastern anomalous arm has only one relatively bright region along
its length. This region is located $\sim$3 times further along the arm from the center of NGC 4258 than 
the bright region in the northwestern arm before bifurcation. The flux density of the intervening 
length of the southeastern arm dims significantly from the central region to a minimum of $\sim$0.05 mJy beam$^{-1}$
at 6 cm and $\sim$0.15 mJy beam$^{-1}$ at 20 cm, and 
then increases to a maximum of 0.42 mJy beam$^{-1}$ and 0.86 mJy beam$^{-1}$ at 6 and 20 cm, respectively.
The beam-corrected, half-power width of the southeastern arm at the position of maximum flux density is
$6{\farcs}7 \pm 1\arcsec$.

Emission (at the $>3\sigma$ level) is detected from the 20 cm radio ``plateaus'' (see Figure 6a)
south of the northwestern arm and north
of the southeastern arm. We detect both broad, diffuse emission and a few large 
structures ($>5\sigma$ level) in the western plateau, while very little diffuse emission is evident from the eastern
plateau. The eastern plateau does show several structures ($>3\sigma$ level) extending into it from the southeastern arm. Since the
radio emission is largely nonthermal (see \S 4.4.2 below), the plateaus 
and any substructuring are much fainter at 6 cm (see Figure 6b).

While much of the plateau emission is faint in these high resolution maps, 
comparison of our lower resolution maps (not shown) with those of others (Hummel et al.~1989; van Albada \& van
der Hulst 1982) reveals more distinct diffuse emission and
extensions from the arms into the plateaus at both wavelengths.
Also, the absolute flux level of the plateaus is underrepresented at 6 cm, and less so at 20 cm, due to missing
very short spacings in the UV coverage. While D configuration observations and zero-spacing fluxes were included in the
imaging process, a shallow ($\sim$1--2$\sigma$), negative ``bowl'' is still present in the 6 cm image, and a shallower one
in the 20 cm image, further hindering the detection of fainter emission in the radio plateaus.

\subsubsection{Spectral Index Map}

Hummel et al.~(1989) determined the spectral index distribution of the radio 
arms at 14$\arcsec$ resolution, and de Bruyn (1977) and Krause et al.~(1984)
at even poorer resolution.
No large-scale spectral index variations were observed, although Hummel et al.~(1989) detected a possible
flattening of the spectrum from $\alpha = -0.7$ to $-$0.5 along the northern ridge of the northwestern arm. We have
generated a spectral index map of the radio arms from our MEM images
to search for variations that may be more evident at higher resolution and sensitivity.

In order to make the spectral index map, we first adjusted the absolute flux density level of the images
as corrections for the negative ``bowls'' discussed above in \S 4.4.1. The background level was sampled along the relatively 
emission-free regions below the southeastern arm and above the northwestern arm
and
determined to be $-$20 $\pm$ 10$\mu$Jy beam$^{-1}$ at 6 cm and $-$15 $\pm$ 10
$\mu$Jy beam$^{-1}$ at 20 cm.  These mean values were subtracted from the corresponding
images. Then, the AIPS task COMB was used to create the spectral index map from the bowl-corrected images. We
applied a clipping level of 90$\mu$Jy beam$^{-1}$ (4.5$\sigma$ at 6 cm, 3$\sigma$ at 20 cm) in order to blank out regions
of low signal-to-noise ratio which tend to give unrealistic spectral index values. 
The resulting spectral index map is shown in Figure 7. 

The nucleus and H~II region candidates are obvious flat spectrum sources 
with $\alpha$ between +0.05 and $-$0.25 (see Table 1), while
the radio arms have a generally
steep spectrum with $\alpha = -0.65 \pm 0.10$. No large-scale spectral index 
variations are apparent along either arm, in agreement with Krause et 
al.~(1984). In addition to the spectral index
map, in Figure 8 we present mean values of $\alpha$ measured using rectangular
apertures typically $8\arcsec \times 8\arcsec$ in size
at a number of 
positions along the anomalous radio arms, in order
to further investigate possible variations not apparent in Figure 7. 
The uncertainties reflect both the standard deviation in the spectral
index measurements within each aperture and uncertainty in the bowl correction.  
The points plotted in Figure 8 are derived from the highest radio surface brightness
regions of the anomalous arms measured from the nucleus outward.
The spectral indices for the 
northwestern and southeastern arms are quite similar; however,
the spectrum of the northern and southern subsections of the northwestern arm 
appears to flatten after bifurcation from $\sim -0.7$ to $\sim -0.3$.  No such 
trend is seen for the southeastern arm.

Note that we are not able to study the
spectral index distribution of the radio plateaus, due to poor signal-to-noise in those regions.
Using $\sim$1$\arcmin$ resolution images at 0.6 and 1.4 GHz, de Bruyn (1977) determined $\alpha$ to be $-$0.7, 
consistent with the arms.  For the integrated radio emission, Krause et al.~(1984) and Hummel et al.~(1989)
report a spectral index of $\alpha = -0.60 \pm 0.03$
for frequencies below 5 GHz, in agreement with our value for the arms, and $-1.16 \pm 0.05$ 
at higher frequencies.

We do not see in either Figure 7 or 8 any significant variation of spectral 
index {\it across\/} the anomalous arms, as claimed to be the case by Hummel
et al.~(1989).  Such a variation is used by Cox \& Downes (1996) as an argument
that the anomalous arms are shocked gas flow due to a bar potential.  Any
variations are on a small spatial scale without any systematic trends.  We see
nothing in our maps to contradict the contention that the arms are associated
with changing jet activity from the nucleus (Cecil et al.~2000) and are in the 
galactic plane.  Cecil et al.~(2000) claim that the jet has recently moved out
of the plane, but above we have already called into question the identity of
their putative radio jet hot spots.

\section{Conclusions}

We have mapped the radio structure of the galaxy
NGC 4258 at 6 and 20 cm with higher resolution than previously available to
identify H~II region and SNR candidates and to determine the
structure of its peculiar anomalous radio arms.  Seven thermal and
five nonthermal radio sources are detected on our $3{\farcs}5$
resolution radio maps
that we identify as H~II region and SNR candidates, respectively. 
We have identified optical counterparts to all of the H~II region candidates
and to three of the SNR candidates
on the H$\alpha$ image by Dutil \& Roy (1999). Two of the SNR candidates also coincide
with X-ray sources (Vogler \& Pietsch~1999).

By comparing the radio and optical measurements of 
the H~II region candidates for the first time,
we obtain an average value of $A_{V}$ = 2.5 $\pm$ 1.0 for the visual extinction toward the
H~II regions
in NGC 4258.
This
extinction in NGC 4258 is similar to that found for M51 and $\sim$1 mag higher than that found for M81 and NGC 6946.
We obtain for NGC 4258 higher values of
extinction from our radio-to-H$\alpha$ method than those obtained using the Balmer
decrement (Dutil 1998), which may be
due to a patchy dust environment within or near the H~II regions. 

We find the bright end of the H~II region radio luminosity function to be fit well
by a power law with an index of $\beta = -1.3 \pm 0.2$ 
at both 6 and 20 cm. 
This value is within the range of indices
derived for other nearby spiral galaxies.

The SNRs are found to be very luminous, comparable to the extraordinary SNR in NGC 6946
(Van Dyk et al.~1994).  Additionally, the LINER nucleus is found to be variable in the
radio, consistent with the behavior of M81's nucleus (Ho et al.~1999).

The anomalous radio arms are found to have no large scale variation in spectral index, 
in agreement with Krause et al.~(1984), although a possible flattening
of the spectral index along the northwestern anomalous sections may be seen. We obtain a mean value of
$\alpha = -0.65 \pm 0.10$ for the arms,
consistent with that found in lower resolution studies. 

\acknowledgements

The authors thank Y.~Dutil and J.-R.~Roy for kindly providing their H$\alpha$ image of NGC 4258 and for helpful
discussions.
CKL thanks the National Research Council and CKL and KWW thank the Office of Naval Research (ONR) for the 6.1
funding supporting this research. 
SDH thanks C.~Paolicchi for her assistance measuring optical sources, and SDH also thanks T.~Loftus of Sweet Briar College for his assistance in obtaining funding for this research. SDH is
supported by
the National
Science Foundation Research in Undergraduate Institutions
Program (grant AST-9970868), the Jeffress Memorial Trust, and funding through Sweet Briar College Faculty Grants.  This research has made use of the NASA/IPAC Extragalactic Database (NED) which is 
operated by the Jet Propulsion Laboratory, California Institute of Technology, under contract with 
the National Aeronautics and Space Administration.

\newpage

\begin{deluxetable}{lccccccccl}
\def\taba{\tablenotemark{a}}
\def\tabb{\tablenotemark{b}}
\def\tabc{\tablenotemark{c}}
\def\tabd{\tablenotemark{d}}
\scriptsize
\def\farcc{\kern 0.08ex\hbox{$.\!\!\phm{^{\prime\prime}}$}\kern -0.08ex}
\def\fsss{\hbox{$.\!\!\phm{^{\rm s}}$}}
\tablenum{1}
\tablecolumns{10}
\tablewidth{0pc}
\tablecaption{Discrete Radio Sources in NGC 4258\taba}
\tablehead{
\colhead{Source}  & \colhead{R.A.(J2000)} & \colhead{Dec.(J2000)} & \colhead{$S_{6}$} 
 & \colhead{$S_{20}$}  & \colhead{$\alpha^{6}_{20}$} & \colhead{$S$(H$\alpha$)\tabb} 
 & \colhead{$A_{V}$(6)} & \colhead{$A_{V}$(20)} & \colhead{$A_{V}$(Bal)\tabc}  \nl
\colhead{} & \colhead{(h m s)}  & \colhead{($\arcdeg$ $\arcmin$ $\arcsec$)}  & \colhead{(mJy)} 
& \colhead{(mJy)} & \colhead{} &
\colhead{} &  \colhead{(mag)} & \colhead {(mag)} & \colhead{(mag)} }
\startdata
1 & 12 18 55.2 & 47 20 35.5 & 1.32 $\pm$ 0.08 & 1.24 $\pm$ 0.15 & +0.05 $\pm$ 0.11 & 2.42 $\pm$ 0.01 & 2.2 $\pm$ 0.1 & 2.0 $\pm$ 0.2
 & 1.81, 1.48 \\ 
2 & 12 18 55.8  & 47 20 24.3 & 0.88 $\pm$ 0.08 & 1.00 $\pm$ 0.15 & $-$0.11 $\pm$ 0.15 & 2.14 $\pm$ 0.01 & 1.8 $\pm$ 0.1 & 1.8 $\pm$ 0.2 & 1.65 \\ 
3 & 12 18 56.5  & 47 20 14.4 & 0.64 $\pm$ 0.11 & 0.81 $\pm$ 0.13 & $-$0.20 $\pm$ 0.20 & 1.36 $\pm$ 0.01 & 2.0 $\pm$ 0.2 & 2.2 $\pm$ 0.2
& 1.98 \\ 
4 & 12 18 57.4  &  47 20 04.6 & 0.50 $\pm$ 0.08 & 0.66 $\pm$ 0.11 & $-$0.22 $\pm$ 0.19 & 1.69 $\pm$ 0.01 & 1.4 $\pm$ 0.2 & 1.6 $\pm$ 0.2
 & 1.33, 1.33 \\ 
5 & 12 18 58.2 & 47 19 03.9 & 0.56 $\pm$ 0.12 & 0.66 $\pm$ 0.18 & $-$0.13 $\pm$ 0.28 & \nodata & \nodata   & \nodata & \nodata \\ 
6  & 12 18 58.2 & 47 19 10.1 & 0.33 $\pm$ 0.09 & 0.33 $\pm$ 0.18 & +0.02 $\pm$ 0.51 &\nodata &\nodata &\nodata &\nodata \\ 
5+6\tabd &\nodata &\nodata & 1.18 $\pm$ 0.12  & 1.65 $\pm$ 0.28  & $-$0.26 $\pm$ 0.13  & 0.52 $\pm$ 0.01 & 4.2 $\pm$ 0.1 & 4.5
 $\pm$ 0.2 & 2.26 \\
7 & 12 18 57.0 & 47 19 03.9 & 0.26 $\pm$ 0.08 & 0.71 $\pm$ 0.14 & $-$0.82 $\pm$ 0.30 &\nodata &\nodata &\nodata  &\nodata \\ 
8  & 12 18 47.9 & 47 18 03.3 & 0.23 $\pm$ 0.06 & 0.43 $\pm$ 0.12 & $-$0.50 $\pm$ 0.32 &\nodata &\nodata & \nodata& \nodata \\ 
9 & 12 19 01.3 & 47 15 25.0 & 1.89 $\pm$ 0.15 & 2.56 $\pm$ 0.22 & $-$0.25 $\pm$ 0.10 & 2.58 $\pm$ 0.02 & 2.6 $\pm$ 0.1 & 2.9 $\pm$ 0.1
 & 1.42 \\ 
10 & 12 18 55.3  & 47 16 47.9  & 0.20 $\pm$ 0.10 & 0.31 $\pm$ 0.07 & $-$0.37 $\pm$ 0.46 & 0.29 $\pm$ 0.01 &\nodata &\nodata  &\nodata \\ 
11  & 12 18 57.6 & 47 16 07.0 & 0.27 $\pm$ 0.09 & 0.53 $\pm$ 0.08 & $-$0.56 $\pm$ 0.31 & 0.15 $\pm$ 0.01 &\nodata &\nodata  &\nodata  \\ 
12 & 12 18 56.3  & 47 16 50.3 & 0.31 $\pm$ 0.09 & 0.64 $\pm$ 0.12 & $-$0.58 $\pm$ 0.29 & 0.55 $\pm$ 0.01 &\nodata & \nodata &\nodata \\ 
nucleus & 12 18 57.5 & 47 18 14.5 & 2.36 $\pm$ 0.16 & 3.14 $\pm$ 0.29 & $-$0.24 $\pm$ 0.10 & 0.46 $\pm$ 0.01  &\nodata &\nodata &\nodata  \\ 
\enddata
\tablenotetext{a}{The optical data are derived from the H$\alpha$ image by Dutil \& Roy (1999).}
\tablenotetext{b}{In units of $10^{-13}$ ergs cm$^{-2}$ s$^{-1}$.}
\tablenotetext{c}{Extinctions determined from Balmer decrements are from Dutil (1998). For apparently single radio sources
corresponding to two optical sources, two extinction values are listed.}
\tablenotetext{d}{The H$\alpha$ flux of a single optical source overlapping sources~5 and 6 was measured. See text in \S 4.1.1} 
\end{deluxetable}

\newpage

\begin{deluxetable}{lccccc}
\scriptsize
\def\taba{\tablenotemark{a}}
\def\tabb{\tablenotemark{b}}
\def\farcc{\kern 0.08ex\hbox{$.\!\!\phm{^{\prime\prime}}$}\kern -0.08ex}
\def\fsss{\hbox{$.\!\!\phm{^{\rm s}}$}}
\tablenum{2}
\tablecolumns{6}
\tablecaption{Background Radio Sources}
\tablehead{
\colhead{Source}  & \colhead{R.A.(J2000)} & \colhead{Dec.(J2000)} &
\colhead{$S_{6}$} 
 & \colhead{$S_{20}$}  & \colhead{$\alpha^{6}_{20}$}  
    \nl
\colhead{}  & \colhead{(h m s)} & \colhead{($\arcdeg$ $\arcmin$ $\arcsec$)} & \colhead{(mJy)} 
& \colhead{(mJy)} & \colhead{} }
\startdata
13 & 12 18 56.5  & 47 23 53.4 & 0.67 $\pm$ 0.10 & 1.81 $\pm$ 0.09 & $-$0.82 $\pm$ 0.13  \\ 
14 & 12 18 46.1  & 47 23 12.4 & 0.36 $\pm$ 0.10 & 0.83 $\pm$ 0.12 & $-$0.70 $\pm$ 0.25  \\ 
15 & 12 19 19.0  & 47 19 07.6  & 0.92 $\pm$ 0.12 & 2.09 $\pm$ 0.23 & $-$0.69 $\pm$ 0.14  \\ 
16a & 12 19 32.6 & 47 18 20.5 & 2.43 $\pm$ 0.20 & 7.76 $\pm$ 0.40 & $-$0.97 $\pm$ 0.08 \\ 
16b & 12 19 32.0 & 47 17 55.8 & 1.75 $\pm$ 0.14 & 2.09 $\pm$ 0.23 & $-$0.15 $\pm$ 0.11  \\ 
16c & 12 19 31.5 & 47 17 48.4 & 0.66 $\pm$ 0.20 & 2.48 $\pm$ 0.11 & $-$1.09 $\pm$ 0.26  \\ 
16d & 12 19 31.0 & 47 17 38.5 & \nodata\taba  & 3.16 $\pm$ 0.16 & \nodata   \\ 
17a & 12 18 46.6  &  47 14 44.0 & 1.42 $\pm$ 0.15 & 5.17 $\pm$ 0.35 & $-$1.07 $\pm$ 0.10  \\ 
17b & 12 18 46.4 & 47 14 27.9  & 0.67 $\pm$ 0.17 & \nodata\tabb  & \nodata    \\ 
17c & 12 18 46.2 & 47 14 19.3 & 3.06 $\pm$ 0.22 & \nodata\tabb  & \nodata   \\
17b+c &\nodata &\nodata & 3.73 $\pm$ 0.28 & 11.68 $\pm$ 0.34 & $-$0.95 $\pm$ 0.07  \\         
\enddata
\tablenotetext{a}{Source 16d is not detected at 6 cm, probably due to primary beam attenuation.}
\tablenotetext{b}{Source 17b cannot be separated from 17c at 20 cm, due to bandwidth smearing. The combined flux
at each wavelength was used to calculate an average spectral index given below for source ``17b+c''.}

\end{deluxetable}

\newpage

\begin{deluxetable}{lcccccc}
\scriptsize
\def\taba{\tablenotemark{a}}
\def\farcc{\kern 0.08ex\hbox{$.\!\!\phm{^{\prime\prime}}$}\kern -0.08ex}
\def\fsss{\hbox{$.\!\!\phm{^{\rm s}}$}}
\tablenum{3}
\tablecolumns{7}
\tablecaption{Star Formation Properties\taba}
\tablehead{
\colhead{Source}  & \colhead{$S^{\rm th}_{6}$} & \colhead{$S^{\rm th}_{20}$} &
\colhead{$L_{\rm T}$} 
 & \colhead{$SFR$}  & \colhead{$N_{\rm UV}$} & \colhead{$\nu_{\rm SN}$}  
    \nl
\colhead{}  & \colhead{(mJy)} & \colhead{(mJy)} & \colhead{(W Hz$^{-1} \times 10^{18}$) } 
& \colhead{($M_{\sun}$ yr$^{-1}$)} & \colhead{(s$^{-1} \times 10^{51}$)} & \colhead{(yr$^{-1} \times 10^{-4}$)} }
\startdata
1 & 1.32 & 1.24 & 8.44 & 0.018 & 6.29 & 7.37 \\     
2 & 0.88 & 1.00 & 5.61 & 0.012 & 4.18 & 4.90     \\
3 & 0.58 & 0.65 & 3.70 & 0.008 & 2.76 & 3.23     \\
4 & 0.45 & 0.50 & 2.84 & 0.006 & 2.12 & 2.45     \\
5 & 0.55 & 0.62 & 3.49 & 0.007 & 2.61 & 3.05     \\
6 & 0.33 & 0.33 & 2.14 & 0.005 & 1.59 & 1.87     \\
9 & 1.60 & 1.80 & 10.2 & 0.022 & 7.61 & 8.91 \\
\enddata
\tablenotetext{a}{Star formation properties were calculated based on $S^{\rm th}_{6}$ and equations in Condon (1992). For sources
with $\alpha < -0.1$, the flux densities given in Table~1 (S$_{6}$ and S$_{20}$)
are corrected for residual nonthermal contamination (\S 4.1.3) to obtain the thermal components
($S^{\rm th}_{6}$ and $S^{\rm th}_{20}$) listed here. For
sources with $\alpha > -0.1$, $S^{\rm th}_{6}$ and $S^{\rm th}_{20}$ are set equal to the values of S$_{6}$ and S$_{20}$.}

\end{deluxetable}

\newpage

\begin{figure}
\figurenum{1}
\caption{({\it a}) 
20 cm $3{\farcs}5$ resolution image of NGC 4258 made using the Maximum Entropy Method.
The labels refer to sources discussed in \S 4.1.1 and 4.1.2.
}
\end{figure}


\begin{figure}
\figurenum{1}
\caption{
({\it b}) 6 cm $3{\farcs}5$ resolution image of NGC 4258 made using the Maximum
Entropy Method.}
\end{figure}


\begin{figure}
\figurenum{2}
\caption{Contour map of the H$\alpha$ image from Dutil \& Roy (1999) overlaid on 
our 20 cm $3{\farcs}5$ resolution greyscale image of NGC 4258 (from Figure 1a). 
The H$\alpha$ contour levels are
7.5$\times$10$^{-18}$ $\times$ (3, 5, 10, 25, 50, 175, 300, 500) ergs cm$^{-2}$ s$^{-1}$.}
\end{figure}

\newpage

\begin{figure}
\figurenum{3}
\plotone{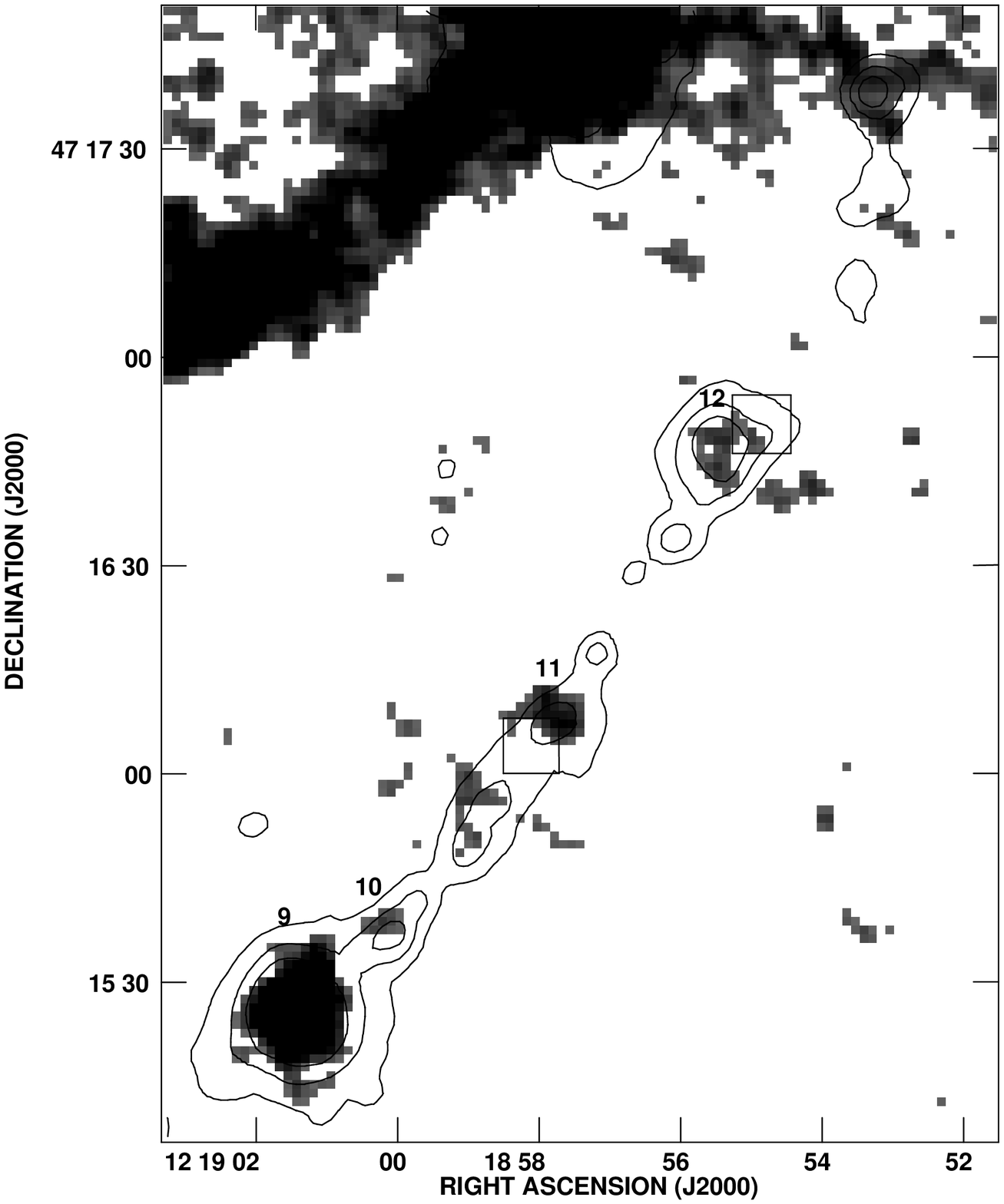}
\caption{Contour map of the southern arm region of the H$\alpha$ image from
Dutil \& Roy (1999) overlaid on the corresponding region of our 20 cm
$3{\farcs}5$ resolution greyscale
image of NGC 4258 (from Figure 1a). Note that all pixels with flux densities less than 90
mJy beam$^{-1}$ (3$\sigma$) have been set to zero in order to better depict the
radio source emission in the southern arm.
The numbers on the figure correspond to sources discussed in \S 4.1.1. The
centers and sizes of the boxes near sources 11 and 12 indicate, respectively,
the positions and positional
uncertainties of two X-ray sources detected by Vogler \& Pietsch (1999).
 The H$\alpha$ contour levels are 7.5$\times$10$^{-18}$ $\times$ (20, 50, 100)
ergs cm$^{-2}$
s$^{-1}$.}
\end{figure}

\newpage

\begin{figure}
\figurenum{4}
\plotone{hymans.fig4.ps}
\caption{Cumulative H~II region luminosity functions at 20 cm
({\it circles}) and 6 cm ({\it diamonds}) for NGC 4258.  Shown are fits to the bright ends of the 
luminosity functions at 20 cm ({\it solid line}) and 6 cm ({\it dashed line}), and have power-law index 
$\beta = -1.3 \pm 0.2$ at both wavelengths.}
\end{figure}

\newpage

\begin{figure}
\figurenum{5}
\plotone{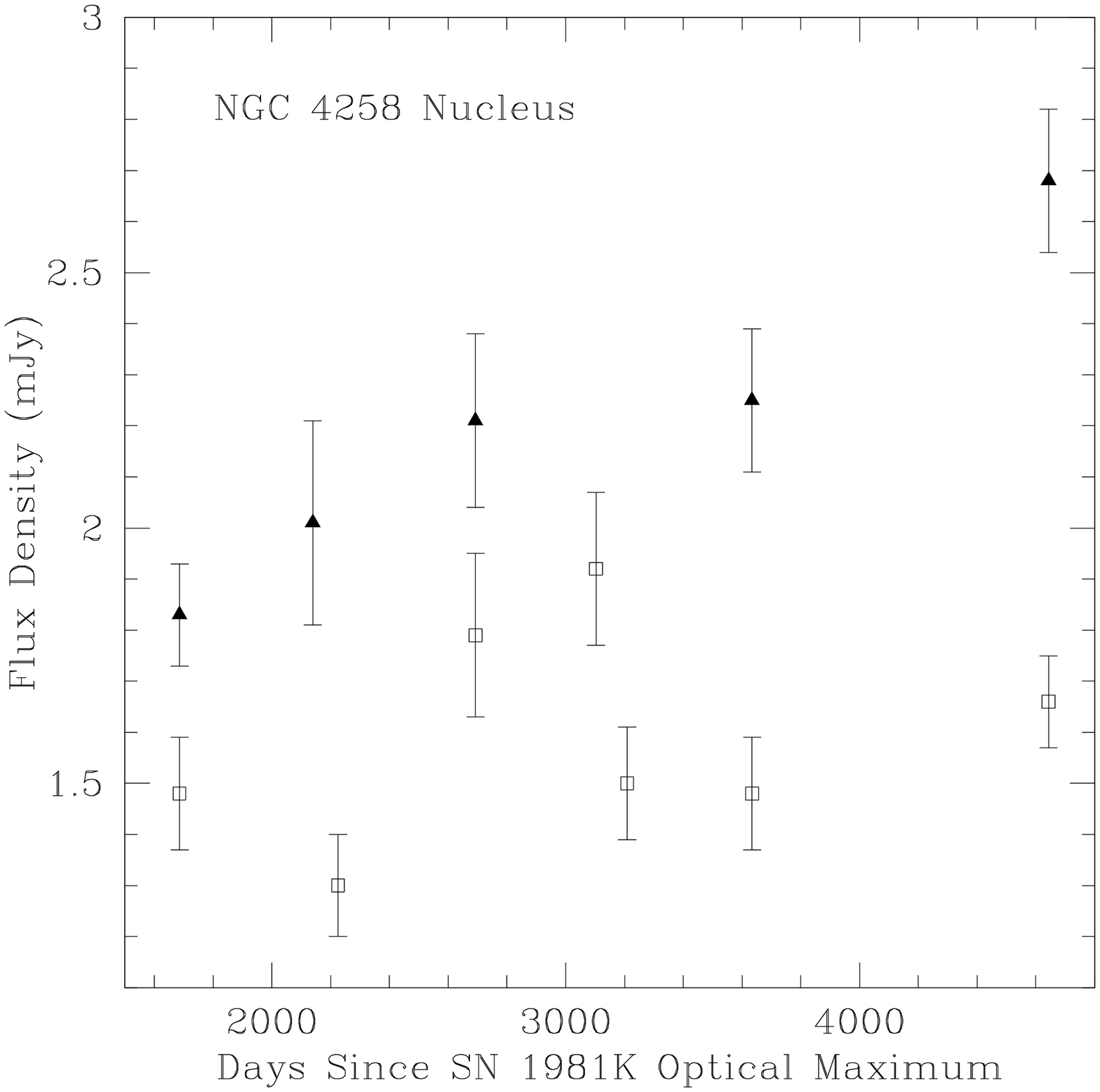}
\caption{Flux density at 6 cm ({\it open squares\/}) and 20 cm ({\it filled
triangles}) of the nucleus of NGC 4258, derived from A-configuration (highest
resolution) VLA
observations, as a function of time (relative to the optical maximum of
SN 1981K in the galaxy; see Van Dyk et al.~1992).  The plot is suggestive of
long-term variability from the nuclear source.}
\end{figure} 

\newpage

\begin{figure}
\figurenum{6}
\caption{({\it a}) 20 cm $3{\farcs}5$ resolution contour map of NGC 4258, emphasizing the emission
from the radio plateaus. The contour levels are 30 $\mu$Jy beam$^{-1}$ $\times$ (3, 5, 7, 9, 11, 13, 15, 17, 19, 21,
23, 25, 27, 29, 35, 40, 45, 50).
}
\end{figure}


\begin{figure}
\figurenum{6}
\caption{
({\it b}) 6 cm $3{\farcs}5$ resolution contour map of NGC 4258. Due to its nonthermal spectrum,
less of the broad disk
emission is detected. The contour levels are 20 $\mu$Jy beam$^{-1}$ $\times$ (3, 5, 7, 9, 11, 13, 15, 17, 19, 21,
23, 25, 27, 29, 35, 40, 45, 50).}
\end{figure}


\begin{figure}
\figurenum{7}
\caption{Map of the spectral index $\alpha$ ($S \propto \nu^{+\alpha}$) between 6 and 20 cm
for NGC 4258. The range of spectral index values are
approximately indicated by the greyscale bar shown at the top of the figure.}
\end{figure}

\newpage

\begin{figure}
\figurenum{8}
\plotfiddle{hymans.fig8.ps}{400pt}{90}{100}{100}{400}{20}
\caption{Spectral index distribution along the anomalous radio arms. The 
{\it circles\/} are samples of the southeastern arm (negative distances). 
The {\it squares\/} represent the northwestern arm before bifurcation and the
southern subsection of the northwestern arm after bifurcation;
the {\it diamonds\/} represent samples of the northern subsection of the northwestern arm after 
bifurcation (positive distances). The
distribution is essentially uniform, although the northwestern arm may
flatten slightly after bifurcation. The flat spectrum point at $\alpha = -0.22$ corresponds
to the nucleus.}
\end{figure}


\begin{thebibliography}{}

\bibitem[Belley \& Roy 1992]{bel92} Belley, J., \& Roy, J.-R. 1992, \apjs, 78, 61.

\bibitem[Blair, W. P.; Fesen, R. A.; Schlegel 2001]{bla01} Blair, W.P., Fesen, R.A., \& Schlegel, 
E.M. 2001, \aj, in press.

\bibitem[Braun \& Walterbos 1993]{bra93} Braun, R., \& Walterbos, R.
1993,
\aaps, 98, 327.

\bibitem[Bridle \& Schwab 1988]{bri88} Bridle, A.H. \& Schwab, F.R. 1988, in ASP
Conf. Ser. 6, Synthesis Imaging in Radio Astronomy, ed. R.A. Perley, F.R. Schwab, \& A.H. Bridle
(San Francisco: ASP), 247.

\bibitem[Cecil et al.~1992]{cec92} Cecil, G., Wilson, A.S., \& Tully, R.B. 1992,
\apj, 390, 365.

\bibitem[Cecil et al.~2000]{cec00} Cecil, G., Greenhill, L.J., DePree, C.G., Nagar, N., Wilson, A.S.,
Dopita, M.A., Perez-Fournon, I., Argon, A.L., \& Moran, J.M. 2000, \apj, 536, 675.

\bibitem[Chary et al.~2000]{cha00} Chary, R., Becklin, E.E., Evans, A.S., Neugebauer, G., Scoville, N.Z.,
Matthews, K., \& Ressler, M.E. 2000, \apj, 531, 756.

\bibitem[Condon 1992]{con92} Condon, J.J. 1992, \araa, 30, 575.


\bibitem[Courtes \& Cruvellier 1961]{cou61} Courtes, G., \& Cruvellier, P. 1961,
CR Acad. Sci. Paris, 253, 218.

\bibitem[Courtes et al.~1993]{cou93} Courtes, G., Petit, H., Hua, C.T.,
Martin, P., Blecha, A., Huguenin, D., \& Golay, M. 1993, \aap, 268, 419.

\bibitem[Cox \& Downes 1996]{cox96} Cox, P., \& Downes, D. 1996, \apj, 473, 219.

\bibitem[de Bruyn 1977]{deb77} de Bruyn, A.G. 1977, \aap, 58, 221.


\bibitem[Dunne, Gruendl, \& Chu 2000]{dun00} Dunne, B.C., Gruendl, R.A., Chu, Y.-H. 2000, \aj, 119, 1172.

\bibitem[Duric et al.~1993]{dur93} Duric, N., Viallefond, F., Goss, W.M.,
\& van der Hulst, J.M. 1993, \aaps, 99, 217.

\bibitem[Duric et al.~1995]{dur95} Duric, N., Gordon, S.M., Goss, W.M.,
Viallefond, F., \& Lacey, C. 1995, \apj, 445, 173.

\bibitem[Dutil 1998]{dut98} Dutil, Y. 1998, PhD Thesis.

\bibitem[Dutil \& Roy 1999]{dut99} Dutil, Y., \& Roy, J.-R. 1999, \apj, 516, 62.

\bibitem[Feinstein 1997]{fei97} Feinstein, C. 1997, \apjs, 112, 29.

\bibitem[Ford et al.~1986]{for86} Ford, H.C., Dahari, O., Jacoby, G.H., Crane, P.C.,
\& Ciardullo, R. 1986, \apj, 311, L7.

\bibitem[Gebel 1968]{geb68} Gebel, W.L. 1968, \apj, 153, 743.


\bibitem[Gordon et al.~1999]{gor99} Gordon, S.M., Duric, N., Kirshner, R.P.,
Goss, W.M., \& Viallefond, F. 1999, \apjs, 120, 247.

\bibitem[Greenhill et al.~1995]{gre95} Greenhill, L.J., Jiang, D.R., Moran, J.M.,
Reid, M.J., Lo, K.Y., \& Claussen, M.J. 1995, \apj, 440, 619.

\bibitem[Herrnstein et al.~1997]{her97} Herrnstein, J.R., Moran, J.M., Greenhill, L.J. Inoue, M,
Nakai, N., Miyoshi, M., \& Diamond, P.J. 1997, BAAS, 191, 25.07.

\bibitem[Ho et al.~1999]{hol99} Ho, L.C., Van Dyk, S.D., Pooley, G.G., Sramek, R.A., \& Weiler, K.W.
1999, \aj, 118, 843.

\bibitem[Hummel et al.~1989]{hum89} Hummel, E., Krause, M., \& Lesch, H. 1989, \aap, 211, 266.

\bibitem[Hyman et al.~2000]{hym00} Hyman, S.D., Lacey, C.K., Weiler, K.W., \& Van Dyk, S.D. 2000, \aj,
119, 1711.

\bibitem[Israel 1980]{isr80} Israel, F.P. 1980, \aap, 90, 246.

\bibitem[Kaufman et al.~1987]{kau87} Kaufman, M., Bash, F.N., Kennicutt,
R.C., \&
Hodge, P.W. 1987, \apj, 319, 61.

\bibitem[Kennicutt, Edgar, \& Hodge 1989]{ken89} Kennicutt, R.C., Edgar,
B.K, \& Hodge, P.W. 1989, \apj, 337, 761.

\bibitem[Krause et al.~1984]{kra84} Krause, M., Beck, R., \& Klein, U. 1984, \aap, 138,385.

\bibitem[Krause et al.~1990]{kra90} Krause, M., Cox, P., Garcia-Barreto, J. A., \& Downes, D.
1990, \aap, 233, L1.


\bibitem[Kronberg, Bierman, \& Schwab 1985]{kro85} Kronberg, P.P.,
Bierman, P.,
\& Schwab, F.R. 1985, \apj, 291, 693.

\bibitem[Lacey, Duric, & Goss 1997]{lac97} Lacey, C., Duric, N., \& Goss, W.M. 1997,
\apjs,
109, 417.

\bibitem[Martin et al.~1989]{mar89} Martin, P., Roy, J.-R., Noreau, L., \& Lo, K.Y. 1989, \apj, 345, 707.

\bibitem[Mathewson et al.~1983]{mat83} Mathewson, D.S., Ford, V.L.,
Dopita,
M.A., Tuohy, I.R., Long, K.S., \& Helfand, D.J. 1983, \apjs, 51, 345.

\bibitem[Miller \& Mathews 1972]{mil72} Miller, J.S. \& Mathews, W.G. 1972,
 \apj, 172, 593.

\bibitem[Miller \& Scalo 1979]{mil79} Miller, G.E., \& Scalo, J.M. 1979, \apjs, 41, 513.

\bibitem[Mills 1983]{mil83} Mills, B.Y. 1983, in Supernova Remnants And
Their
X-Ray Emission, ed.~J.~Danziger, \& P.~Gorenstein, P. 551.

\bibitem[Miyoshi et al.~1995]{miy95} Miyoshi, M., Moran, J., Herrnstein, J., Greenhill, L., Nakai, N.,
Diamond, P., \& Inoue, M. 1995, Nature, 373, 127.

\bibitem[Oort 1987]{oor87} Oort, M.J.A. 1987, \aaps, 71, 221.

\bibitem[Pietsch et al.~1995]{pie94} Pietsch, W., Vogler, A., Kahabka, P., Jain, A.,
\& Klein, U. 1994, \aap, 294, 386.

\bibitem[Rozas, Beckman, \& Knapen 1996]{roz96} Rozas, M., Beckman, J.E.,
\& Knapen, J.H. 1996, \aap, 307, 735.

\bibitem[Sandage \& Tammann 1981]{san81} Sandage \& Tammann 1981, A
Revised
Shapley-Ames Catalog of Bright Galaxies (Washington: Carnegie Inst. Washington)

\bibitem[Sarazin 1976]{sar76} Sarazin, C.L. 1976, \apj, 208, 323.

\bibitem[Srinivasan et al.~1984]{sri84} Srinivasan, G., Dwarakanath, K.~S., \&
 Bhattacharya, D. 1984, JApA, 5, 403. 

\bibitem[Turner \& Ho]{tur94} Turner, J.L., \& Ho, P.T.P. 1994, \apj, 421, 122.

\bibitem[Ulvestad \& Antonucci]{ulv97} Ulvestad, J. S. \& Antonucci,
R.R.J. 1997, \apj, 488, 621.

\bibitem[van Albada \& van der Hulst 1982]{van82} van Albada, G.D., \& van der Hulst, J.M. 1982,
\aap, 115, 263.

\bibitem[van der Hulst et al.~1988]{van88} van der Hulst, J.M., Kennicutt,
R.C., Crane, P.C., \& Rots, A.H. 1988, \aap, 195, 38.


\bibitem[van der Kruit et al.~1972]{van72} van der Kruit, P.C., Oort, J.H., \& Mathewson, D.S. 1972,
\aap, 21, 169.

\bibitem[Van Dyk et al.~1992]{vand92} Van Dyk, S.D., Weiler, K.W.,
Sramek, R.A., \& Panagia, N. 1992, \apj, 396, 1995.

\bibitem[Van Dyk et al.~1994]{vand94} Van Dyk, S.D., Sramek, R.A., Weiler, K.W., Hyman, S.D., \&
Virden, R.E. 1994, \apjl, 425, L77.

\bibitem[Viallefond \& Goss 1986]{via86} Viallefond, F. \& Goss, W.M.
1986, \aap,
154, 357.

\bibitem[Vogler \& Pietsch 1999]{vog99} Vogler, A., \& Pietsch, W. 1999, \aap, 352, 64.

\bibitem[Weiler \& Panagia 1978]{wei78} Weiler, K.W. \& Panagia, N. 1978,
\aap, 70, 419.

\bibitem[Weiler et al.~1986]{wei86} Weiler, K.W., Sramek, R.A., Panagia,
N.,
van der Hulst, J.M., \& Salvati, M. 1986, \apj, 301, 790.

\end{thebibliography}
\end{document}